\documentclass[pre,aps,twocolumn,preprintnumbers,superscriptaddress]{revtex4}

\usepackage{graphicx}
\usepackage{dcolumn}
\usepackage{bm}
\usepackage{amssymb}
\usepackage{pifont}
\usepackage{amsmath}
\usepackage{times}
\usepackage[nooneline]{subfigure}
\usepackage{hyperref}

\begin{document}
\title{Maximizing the statistical diversity of an ensemble of
bred vectors \\ by using the geometric norm}

\author{Diego Paz\'o}
\affiliation{Instituto de F\'{\i}sica de Cantabria (IFCA), CSIC--Universidad de
Cantabria, E-39005 Santander, Spain}

\author{Miguel A. Rodr{\'\i}guez}
\affiliation{Instituto de F\'{\i}sica de Cantabria (IFCA), CSIC--Universidad de
Cantabria, E-39005 Santander, Spain}

\author{Juan M. L{\'o}pez}
\affiliation{Instituto de F\'{\i}sica de Cantabria (IFCA), CSIC--Universidad de
Cantabria, E-39005 Santander, Spain}

\begin{abstract}
We show that the choice of the norm has a great impact on the construction of 
ensembles of bred vectors. The geometric norm maximizes (in comparison with
other norms like the Euclidean one) the statistical diversity of the
ensemble while, at
the same time, enhances the growth rate of the bred vector and its projection on the linearly most
unstable direction, {\it i.e.}~the Lyapunov vector. 
The geometric norm is also optimal in providing the least fluctuating
ensemble dimension among all the spectrum of $q$-norms studied.
We exemplify our
results with numerical integrations of a toy model of the atmosphere (the
Lorenz-96 model), but our findings are expected to be generic for spatially
extended chaotic systems.
\end{abstract}

\draft{[J.Atmos.Sci.{\bf{68}},1507--1512(2011)]\href{http://www.ametsoc.org/pubs/copyrightinfo/ams_copyright_policy_2010.pdf}{\textcopyright Copyright July 2011 AMS}}

\maketitle

\section{Introduction}
\label{sec_intro}
The `breeding method' is a well-established and computationally inexpensive
procedure for generating perturbations for ensemble
integrations~\citep{toth93,toth97}. Bred vectors (BVs) are finite perturbations
periodically rescaled to a certain magnitude that have been prominently used in
probabilistic weather forecasting with ensembles~\citep{Kalnay,gneiting05}. The
breeding method and variants of it are applied in operative ensemble forecast
systems, such as National Centers for Environmental Predictions (NCEP, USA), see
e.g.~\cite{wei08}. Moreover, breeding continues to be a popular tool to study
the predictability of a variety of systems such as the baroclinic rotating
annulus~\citep{young08} or the atmosphere of
Mars~\citep{newman04}.

Different initial BV perturbations all generally tend to become aligned with the
fastest growing modes. If different BVs were
globally quasi-orthogonal to each other~\citep{toth97}, one might
expect they would automatically provide a good sample of the different dominant
growing error
directions, without the need for additional computation.
A closer inspection reveals  that the BV perturbations are often
locally rather similar in shape, differing only in sign and
amplitude~\citep{toth97,hallerberg10}. In fact, a major modification of the BV
implementation at NCEP has recently been implemented by replacing the BVs given
by the ensemble forecast with some `ensemble transform' that orthogonalizes the
ensemble with respect to the metric defined by the inverse covariance
matrix~\citep{bishop99,wang03,wei06,wei08}. Other metrics can be used and lead
to different ensembles of BVs~\citep{keller10}. Orthogonalization with respect
to a given metric generally enhances the statistical diversity of the ensemble
by making the BV perturbations globally more
dissimilar~\citep{annan04,keller10}.

In this paper we show how the ensemble diversity can be enhanced 
by using the geometric norm with no further transforms or orthogonalizations
needed. We first show that the BVs dynamics and the statistical properties of
the ensemble strongly depend on the norm definition used to construct them. So
far Euclidean-type norms are widely used in applications. However, our results
demonstrate that, among a spectrum of studied norms, the geometric norm is the
most convenient because it provides a greater statistical diversity of the
ensemble, while it enhances the projection of the ensemble as a whole on the
most unstable direction. With other norm choices, like the standard Euclidean
one, a good projection on the leading Lyapunov vector (LV) is always associated
with the collapse of all the BVs, i.e.~the complete loss of the ensemble
diversity.

\section{The model}
We illustrate our study with numerical integrations of the well-known Lorenz-96
model~\citep{lorenz96} that has been used by various authors as a low order
testbed for atmospheric prediction and assimilation studies
\citep{lorenz98,anderson01}. This model is defined by the set of variables
$\{u(x,t)\}_{x=1,\ldots,L}$ and evolves according to
\begin{eqnarray}
\frac{d \,u(x,t)}{dt}&=&- u(x-1,t)\left[u(x-2,t)
 -u(x+1,t)\right] \nonumber\\   &-& u(x,t)+ F,\quad 
\mbox{with}\quad x=1,...,L.
\label{lorenz}
\end{eqnarray}
with periodic boundary conditions in the discrete spatial variable $x$.
Hereafter we adopt a system size of $L=128$ and a forcing constant $F=8$. 
For these values the system exhibits well developed chaos~\citep{lorenz06}.

A good description of the chaotic dynamics can be achieved by
understanding the behavior of initial infinitesimal perturbations, which
are governed by the `tangent linear model':
\begin{eqnarray}
\frac{d \,\delta u(x,t)}{dt}=- \delta u(x-1,t)\left[u(x-2,t)-u(x+1,t)\right] \nonumber\\
- u(x-1,t)\left[\delta u(x-2,t)- \delta u(x+1,t)\right] -\delta u(x,t) .
\label{tangent}
\end{eqnarray}
After some transient any infinitesimal perturbation $\delta u(x,0)$ becomes
permanently aligned along the most unstable direction.
This direction defines, disregarding an arbitrary nonzero constant factor,
the leading LV, and hereafter denoted
$\mathbf{g}(t)=\{g(x,t)\}_{x=1,\ldots,L}$. 

Obtaining the tangent linear (and adjoint) models can be however extremely
difficult in operative weather models and one has to resort to analyzing finite
perturbations, which are evolved with the full nonlinear model. This is for
instance the situation at NCEP, where ensembles of BVs are used.

\section{Bred vectors}
\label{model}

BVs are finite perturbations obtained after periodic
rescaling, say at times $t_m=m T$ ($m\in
\mathbb{Z}^{+}$).
A control trajectory $\mathbf{u}$ and a perturbed
one $\mathbf{u}'$, are simultaneously integrated [via Eq.~(\ref{lorenz})] 
and at the scheduled times
the difference between them is calculated
\begin{equation}
\mathbf{\Delta u}(t_m)= \mathbf{u}'(t_m) - \mathbf{u}(t_m) 
\label{dif}
\end{equation}
and rescaled to a given amplitude $\varepsilon$, obtaining the BV 
\begin{equation}
\mathbf{b}(t_m)=\varepsilon \frac{\mathbf{\Delta u}(t_m)}{\|\mathbf{\Delta
u}(t_m)\|} 
\label{scaling}
\end{equation}
This BV is then used to redefine the perturbed system:
\begin{equation}
\mathbf{u}'(t_m^+)= \mathbf{u}(t_m)+ \mathbf{b}(t_m).
\label{pert}
\end{equation}
with $t_m^+=\lim_{\nu\to 0} t_m +\nu$. The perturbed $\mathbf{u}'$ and control
$\mathbf{u}$ states are then evolved in time according to
the model equations, Eq.~(\ref{lorenz}), until the next scheduled rescaling.
At the next scheduled time $t_{m+1}$ 
the breeding cycle, Eqs.~(\ref{dif})-(\ref{pert}), is repeated.
After several breeding cycles, the perturbations generated by this procedure 
acquire
a large growth rate, which makes them suitable for ensemble forecasting.
Usually a set of BVs is evolved from different initial random perturbations and
this constitutes the ensemble. Ideally a good ensemble of BVs should span the
most unstable directions in phase space well enough to capture the main
instabilities.

There are three basic ingredients in the definition of the BV: (i) the
rescaling interval $T$, (ii) the perturbation amplitude $\varepsilon$, and
(iii) the choice of the norm $\|\cdot\|$ used in Eq.~(\ref{scaling}). 

The rescaling interval $T$
has a negligible influence in the results as long as it remains small
---say, smaller than the doubling time, which is on the order of $0.4$ time
units (t.u.) for the Lorenz-96 model.
We have used $T=0.1$ t.u., which corresponds to $1/2$ day in the time scale
assumed
by~\cite{lorenz96}. 

The perturbation amplitude $\varepsilon$
controls the ``finiteness'' of
the perturbations; a sufficiently small $\varepsilon$
makes the perturbation quasi-infinitesimal, and in the limit
$\varepsilon\to 0$ the BV perfectly aligns
with the leading LV of the system.

However, very little is known about the effect of the norm choice on the
properties of the resulting ensemble and we discuss this
issue in detail in the incoming sections.

\section{Choice of a norm}

The choice of the norm is probably the more obscure element
determining the BVs' nature. BVs have often been claimed to be insensitive
to the choice of norm~\citep{kalnay02,corazza03}.
However, this belief is not actually based on any rigorous argument.
Here we show that the effect of changing the norm type has a dramatic impact on
BVs. We will show that different norms lead to different ensemble properties
and it is not a mere change of the `ruler' or metrics.
There are intrinsic and genuine
effects on the statistics of the BVs for each particular norm type.

Intuitively, for a homogeneous system
like the Lorenz-96 model, any definition for the norm one wants to use should be
homogeneous in the sense that it weights equally all sites.
To see why this constraint is relevant let us 
consider a particularly illustrative example. Think of 
a norm arising from some scalar product
$\|\mathbf{b}\|^2 = \left< \mathbf{b},\mathbf{b}\right> = \mathbf{b}^{\rm T} 
\bm{\mathsf{M}} \mathbf{b}$
with a very ``unbalanced'' metric 
matrix $\bm{\mathsf{M}}$, e.g.~$\bm{\mathsf{M}}=\rm{diag}(100,1,1,\ldots,1)$.
This choice would result in very dissimilar BVs 
depending if the site $x=1$ is more or less unstable at a given time. 
For a given $\varepsilon$,
at some times the vector dynamics could be infinitesimal-like while 
at other moments it
would be clearly finite. 
For spatially homogeneous systems, it is reasonable to restrict ourselves
to ``homogeneous`` norms that produce a BV that
is statistically equivalent up to a high degree at different times and we do so
in our study.

In this work we compare the performance of $q$-norms, which are defined as
\begin{equation}
\|\mathbf{\Delta u}\|_q  = \left[ \frac{1}{L} \sum_{x=1}^L |{\Delta u}(x,t)|^q
\right]^{1/q} 
\end{equation}
Note that for $q=2$ the norm is an energy-like norm, analogous to those used in
atmospheric models.
In the limit $q\to\infty$ the $q$-norm becomes the supremum norm:
\begin{equation}
\|\mathbf{\Delta u}\|_{q\to\infty}=\sup\{|{\Delta u}(x,t)|\}_{x=1,\ldots,L}. 
\end{equation}
Moreover, the geometric mean is obtained in the limit\footnote{$\lim_{q\to0} \|\mathbf{\Delta u}\|_q =
\lim_{q\to 0} \exp[q^{-1} \ln(L^{-1} \sum_{x_=1}^L |\Delta u(x)|^q)]=
\lim_{q\to 0} \exp[q^{-1} \ln(L^{-1} \sum_{x_=1}^L e^{q\ln|\Delta u(x)|})]$ = Eq.~(\ref{0norm}).}
$q\to 0$:
\begin{equation}
\|\mathbf{\Delta u}\|_0  =\prod_{x=1}^L \left|\Delta u(x,t)\right|^{1/L} 
\label{0norm}
\end{equation}
The use of the geometric norm yields the so-called {\em logarithmic} BVs
\citep{primo05,primo06,primo08,pazo10,hallerberg10}. For clarity of presentation
we will add
the subscript $q$ to the notation (for the BV, $\mathbf{b}_q$, and
the amplitude, $\varepsilon_q$) to emphasize which $q$-norm is being used. 
For all $q$-norms the BVs look
like very similar to the naked eye
is strongly localized in space, and are strongly localized in space for small $\varepsilon_q$. 
(see for instance~\cite{hallerberg10}
and~\cite{primo08} for typical snapshots of BVs with $q=0$ and $q=2$).

Along this paper we shall be considering an ensemble of $k=10$ BV
members,
$\{\mathbf{b}_q^{(i)}\}_{i=1,\ldots,k}$. All
members of the ensemble are simultaneously rescaled, and they are initiated
with independent random initial conditions,
which is expected to result in some degree of diversity
in the ensemble of BVs \citep{kalnay02}.

\section{Results}

We define ensemble diversity as the degree of linear independence or
transversality among the ensemble members. Diversity can be quantified by
calculating the `ensemble dimension', which measures the effective dimension of
the sub-space spanned by the ensemble. The higher the ensemble dimension the
greater the statistical diversity. Higher ensemble dimension would imply larger
dissimilarities among the ensemble members. Since
BV perturbations tend to align with the fastest growing modes, a greater
diversity indicates that the ensemble is able to actually sample not only the
main LV but also other, less unstable, directions.

Our goal here is to show, by means of several numerical calculations with a
simple model, that the 0-norm is more convenient than other norms for
constructing ensembles of BVs as far as ensemble diversity enhancement is
concerned. We arrive to this conclusion measuring the ensemble dimension and its
temporal fluctuations, the average growth rate, and the alignment of the
ensemble members with the main LV.

\subsection{Ensemble dimension}

In this subsection we will analyze the statistical diversity in an ensemble of
$k$ BVs. Clearly, for all $q$ values, in the limit $\varepsilon_q \to 0$ all BVs
become aligned with the leading LV and there is no diversity
in the ensemble (other than a global sign for the orientation of the vectors).
If $\varepsilon_q$ is finite some degree of transversality between ensemble
members can be expected, and to measure this diversity of the ensemble we resort
to the so-called ensemble dimension~\citep{bretherton99}.
\begin{figure}[t]
  \noindent\includegraphics[width=19pc,angle=0]{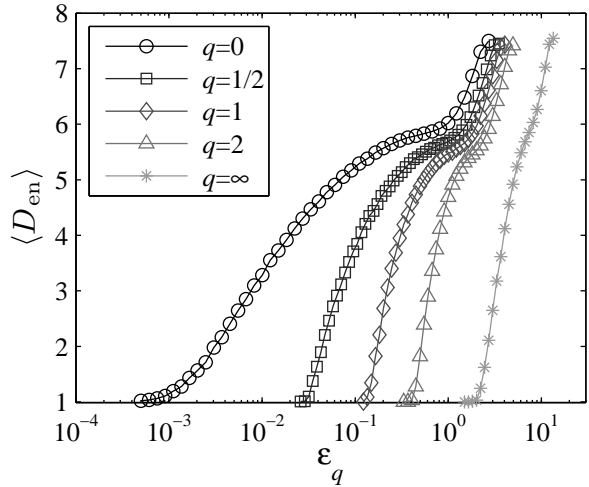}\\
  \caption{Average ensemble dimension as a function of the amplitude $\varepsilon_q$ for
ensembles of $k=10$ BVs, with different $q$-norms.}\label{epsilon}
\end{figure}

The ensemble dimension~\citep{oczkowski05}
or BV dimension~\citep{patil01}
was proposed as a way to account for the number of effective
degrees of freedom that explains most of the total ensemble variance (in the spirit of
principal component analysis), see e.g.~\cite{bretherton99} and references
therein.
To compute the ensemble dimension at a given time
one computes the $k\times k$ covariance matrix $\bm{\mathsf{C}}$ with elements:
\begin{equation}
C_{ij} (t) =\frac{\left<\mathbf{b}_q^{(i)},\mathbf{b}_q^{(j)}\right>}{L
\|\mathbf{b}_q^{(i)}\|_2 \|\mathbf{b}_q^{(j)}\|_2} .
\label{c}
\end{equation}
where the standard scalar product is used in the numerator
$\left<\mathbf{b}_q^{(i)},\mathbf{b}_q^{(j)}\right> =\sum_x b_q^{(i)}(x,t) 
b_q^{(j)}(x,t)$.
If we denote by $\{\mu_i(t)\}_{i=1,\ldots,k}$ the set of eigenvalues of
$\bm{\mathsf{C}}$, the ensemble dimension is:
\begin{equation}
 D_\mathrm{en}(t)=\frac{\left(\sum_{i=1}^k \sqrt{\mu_i} \right)^2}{\sum_{i=1}^k
\mu_i}
\label{den}
\end{equation}
where the denominator ($\sum_i \mu_i$) equals $k$  due to the normalizing terms
in the denominator of (\ref{c}).
The statistic (\ref{den}) typically returns a real number between two limit
values:
$D_\mathrm{en}=k$ (if all vectors are orthogonal)
and $D_\mathrm{en}=1$ (if all vectors are aligned). Therefore $D_\mathrm{en}(t)$
measures the instantaneous degree of ``transversality'' of the ensemble. 
 
Figure~\ref{epsilon} depicts the results of the time-average ensemble dimension
$\left< D_\mathrm{en}\right>$
for different $q$-norms. Depending on the value of $q$ the amplitude
$\varepsilon_q$ is varied in a different range. The largest value of
$\varepsilon_q$ in each data set corresponds (approximately) to the value of the
average distance (for the corresponding $q$-norm) between independent
realizations of the model (i.e.~random climatological values). In applications
$\varepsilon_q$ is much smaller than this value (typically of the order of the
analysis error). In the small $\varepsilon_q$ region of the plots,
$\left< D_\mathrm{en}\right>$ becomes equal to 1 below a particular
value of $\varepsilon_q$, though for the 0-norm the convergence to
$1$ appears to be much less abrupt.

\subsection{Fluctuations of the ensemble dimension}

Figure~\ref{epsilon} shows that all $q$-norms allow
to obtain ensembles with a certain $\left< D_\mathrm{en}\right>$
after tuning $\varepsilon_q$ to a particular value.
However the ensemble dimension is a time-fluctuating quantity
and one should wish to minimize its fluctuations. Of course some degree
of fluctuations is unavoidable due to (i) finiteness of the ensemble
and (ii) intrinsic fluctuations in the state of the system (which progressively
average out for large enough systems).

\begin{figure}[t]
  \noindent\includegraphics[width=19pc,angle=0]{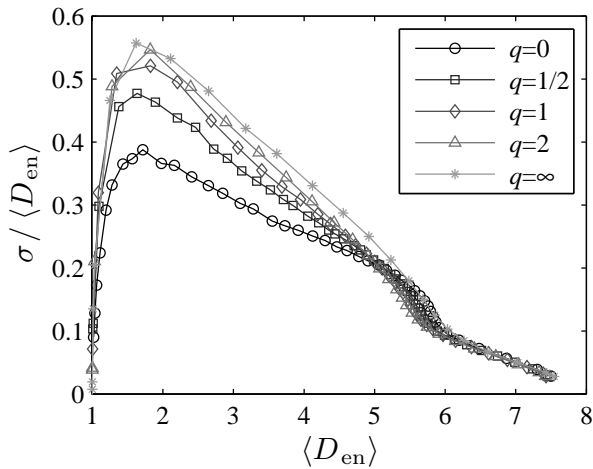}\\
  \caption{Relative fluctuations of the ensemble dimension.}\label{sigma}
\end{figure}

We characterize the fluctuations of $D_\mathrm{en}$ by means of the standard
deviation
$\sigma= \left<\left[D_\mathrm{en}(t)-\left< D_\mathrm{en}\right>\right]^2
\right>^{1/2}$
where the brackets denote a temporal average. The results
are depicted in Fig.~\ref{sigma}, where we plot the relative
fluctuations of the ensemble dimension
versus $\left< D_\mathrm{en}\right>$
to better compare different $q$-norms. 
One can readily see that the 0-norm produces the ensemble with the smallest
fluctuations for most $\left< D_\mathrm{en}\right>$ values.

\subsection{Alignment with the main Lyapunov vector}

\begin{figure}[t]
  \noindent\includegraphics[width=19pc,angle=0]{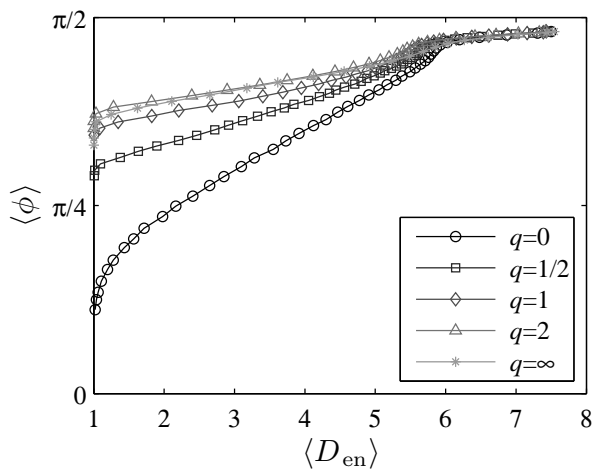}\\
  \caption{Average angle between the BVs and the main LV.}\label{phiq}
\end{figure}

Ideally (i.e.~disregarding limitations by numerical accuracy) the BVs become
perfectly aligned with the main LV as $\varepsilon_q\to 0$. To determine
quantitatively the degree of alignment with the LV, $\mathbf{g}(t)$, we have
measured the instantaneous angle between each BV of the ensemble,
$\mathbf{b}_q^{(i)}(t)$, and $\mathbf{g}(t)$ at breeding times $t=t_m$ as
customary in a $L$-dimensional Euclidean space\footnote{As the sign of the
LV is not defined, we can adopt the convention of defining $\phi$
in the range
$[0,\tfrac{\pi}{2}]$.}:
\begin{equation}
\phi^{(i)}(t)=\measuredangle\left(\mathbf{g}(t=t_m),
\mathbf{b}_q^{(i)}(t=t_m)\right) 
\end{equation}
The ensemble and time average angle $\left< \phi \right>$ is shown in
Fig.~\ref{phiq}, and demonstrates that the logarithmic
BVs ($q=0$)
are able to achieve a considerable degree of alignment with the LV
on average,
while retaining some degree of diversity. 
One clearly sees that BVs constructed with norms $q > 0$ become
strongly aligned among themselves while still
keep a high degree of transversality  with the
main LV, as reflected by the high average angle of the ensembles ($\left< \phi
\right> > \pi/4$) in Fig.~\ref{phiq} for $\left< D_\mathrm{en}\right> = 1$. In
contrast, the `logarithmic ensemble' ($q=0$) exhibits a lower angle with
the main LV, even if the statistical diversity is high. 
We claim that
the higher diversity and lower $\left< \phi \right>$ exhibited by the ensemble
of logarithmic BVs ($q=0$), as compared with the ensembles with $q>0$, indicates
that this ensemble is spanning a
sub-space formed by a narrow hyper-cone around the main LV, while ensembles with
$q>0$ tend to lie in a lower dimension subspace that is more
transverse to the LV.

\subsection{Average growth rate}
Also the growth rate of the ensemble members can be used to 
compare with that of the main LV, reflecting again the different behavior for
different norm choices.
The average exponential growth rate of 
the bred vectors is
\begin{equation}
\lambda=\frac{1}{T}\left< \ln\left( \frac{\|\mathbf{\Delta
u}(t_m+T)\|_2}{\|\mathbf{b}_q(t_m)\|_2}\right) \right>
\end{equation}
Notice that, for the sake of clarity we are using the same norm type ($q=2$) to
measure the exponential growth rate in all cases
(nevertheless due to the long averaging the norm type is irrelevant).

\begin{figure}[t]
  \noindent\includegraphics[width=19pc,angle=0]{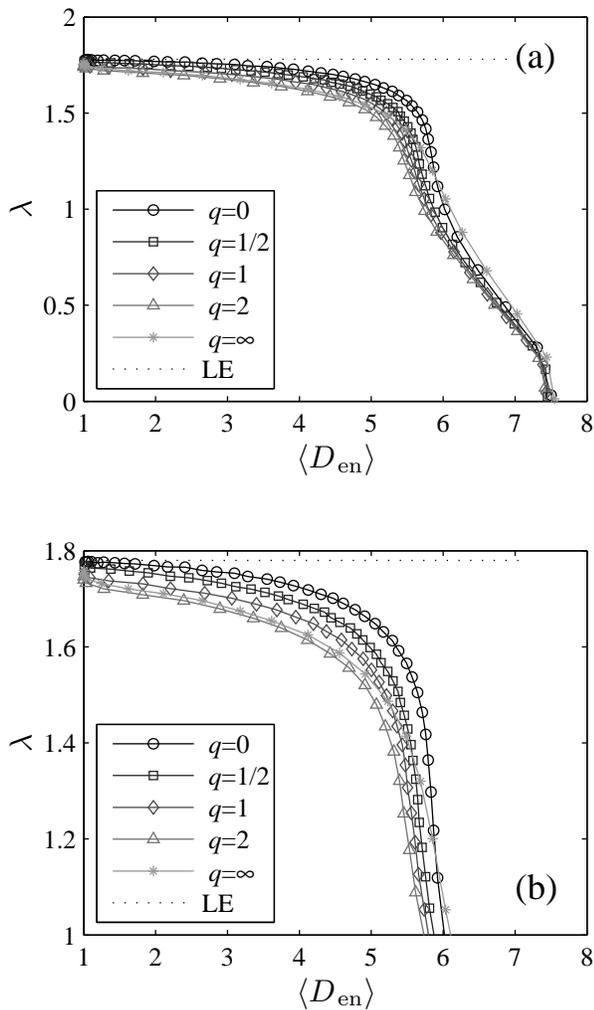}\\
  \caption{(a) Average growth rate $\lambda$ as a function of the average
ensemble dimension.
The dotted line indicates the value of the Lyapunov exponent.
(b) Zoom of panel (a).}\label{leq}
\end{figure}

Figure~\ref{leq} shows the dependence of $\lambda$ on the ensemble dimension.
One can see that the logarithmic BVs ($q=0$) exhibit the largest amplification
rate for a given ensemble dimension, which is in agreement with the results
discussed in the preceding subsections showing that the logarithmic ensemble
($q=0$), among all ensemble choices, exhibited the greatest projection on the
LV. Conversely, given an exponential growth rate $\lambda$, using
the $0$-norm will result in the most diverse ensemble.

\section{Conclusions}
We have studied the effect of different norms on the construction of ensembles
of BVs. The geometric ($q=0$) norm outperforms other norms (like the Euclidean
one, $q=2$) for constructing ensembles of BVs in spatially extended systems.
The enhancement of performance (in terms of root-mean square error,
ensemble spread, and calibration time) of ensembles of logarithmic ($q=0$)
bred vectors 
with respect to standard ``Euclidean'' ($q=2$) bred vectors was already
uncovered by~\cite{primo08}.
In the present work we give a rationale behind those results. We show
that an ensemble of logarithmic BVs
(obtained with the 0-norm) exhibits greater diversity ---larger ensemble
dimension--- while its members are more strongly projected on the leading
LV and have growth rates that rapidly approach the leading
Lyapunov exponent. In comparison, ensembles based on BVs with $q > 0$ perform
rather poorly. They tend to collapse in one single direction (i.e.,~$\left<
D_\mathrm{en}\right> = 1$) very abruptly as the BV amplitude is diminished and,
even when all the statistical diversity is lost, they remain rather
transverse to the
leading LV as demonstrated by the angle with the main LV shown in 
Fig.~\ref{phiq}.
Moreover, the geometric norm also leads to the 
least fluctuating ensemble
dimension among all the possible $q$-norms. 

In the view of these results two prominent questions remain open. On the
one hand, it would be very interesting to evaluate the performance
of $0$-norm BVs in real applications. The study by \cite{primo08} of 0-norm BVs
already showed promissing, albeit preliminary, results.
Clearly, more research is needed in this direction. On the
other hand, there is the 
problem of analyzing the potential advantages of ensemble
Kalman filters based on $0$-norm BVs. Our results show that
logarithmic BVs have very nice properties regarding statistical
diversity, growth rates, and projection onto the main LV. Therefore,
a natural question that arises is: to what extent can these features
translate into a better 
performance of ensemble Kalman filtering methods?. We believe our results
may serve as a basis for future research along these lines.\\

D.P.~acknowledges support by CSIC under the Junta de Ampliaci\'on de
Estudios Programme (JAE-Doc). Financial support from
the Ministerio de Ciencia e Innovaci\'on (Spain) under Projects
No.~FIS2009-12964-C05-05 and No.~CGL2010-21869/CLI is acknowledged.

\end{document}